\documentclass[twocolumn,showpacs,amsmath,amssymb,aps,prd,floatfix,longbibliography]{revtex4-2}  
\usepackage{amssymb,amsmath}
\usepackage{color}
\usepackage{graphicx}
\usepackage{dcolumn}
\usepackage{bm}
\usepackage{upgreek}
\usepackage{marvosym}
\usepackage{latexsym,epsfig} 
\usepackage{bm}
\usepackage{mathtools}
\usepackage[T1]{fontenc}
\usepackage{hyperref}
\usepackage[nodisplayskipstretch]{setspace}

\usepackage{upgreek}
\usepackage{marvosym }

\DeclareUnicodeCharacter{2212}{-}

\begin{document}
\title{Dielectric Screening and Electric Field Control of Ferromagnetism at the CaMnO$_3$/CaRuO$_3$ Interface}  
\author{Churna Bhandari}
\altaffiliation[Current Address: ]
{The Ames Laboratory, U.S. Department of Energy, Iowa State University, Ames, IA
50011, USA}
\author{S. Satpathy}
\affiliation{Department of Physics \& Astronomy, University of Missouri,
Columbia, Missouri 65211, USA}  
%
\begin{abstract}
Control of magnetism by an applied electric field is a desirable technique for the functionalization of magnetic materials.  
Motivated by recent experiments, we study the electric field control of the interfacial magnetism of  
CaRuO$_3$/CaMnO$_3$ (CRO/CMO) (001), 
a prototype interface between a non-magnetic metal and an antiferromagnetic insulator. 
Even without the electric field,
the interfacial CMO layer acquires a  ferromagnetic moment due to a spin-canted state,
caused by the Anderson-Hasegawa double exchange (DEX) 
 between the  Mn moments  and
the 
leaked electrons from the CRO side. 
An electric field would alter the carrier density at the interface, leading to the possibility of controlling the magnetism,
since DEX is sensitive to the carrier density. 
We study this effect quantitatively using
 density-functional calculations in the slab geometry.
 We find a text-book like dielectric screening of the electric field,
which introduces polarization charges at the interfaces and the surfaces. 
The extra charge at the interface enhances the ferromagnetism via the DEX interaction,
while away from the interface the original AFM state of the Mn layers
 remains unchanged. 
The effect could have potential application in spintronics devices.

 %

\end{abstract}  
\date{\today}					
\maketitle


\section { Introduction}


There is a considerable interest in controlling the magnetism of magnetic materials by an external electric field because of its potential 
applications in spintronics.
Heterostructures between transition metal oxides have been identified
as possible platforms for achieving this magnetoelectric coupling effect. \cite{TokuraAPL01,HitoshiSc98,GibertN12,GrutterPRL13} 
One such prototypical interface is the 
 (001) interface  between the paramagnetic metal CaRuO$_3$ (CRO) and the antiferromagnetic insulator CaMnO$_3$ (CMO),
which has been well studied, both experimentally and theoretically. \cite{TokuraAPL01, NandaPRL07, Freeland2010, Suzuki2012, Suzuki2015}
While CMO is an antiferromagnetic insulator in the bulk, the interface layer adjacent to the paramagnetic CRO acquires 
a net ferromagnetic moment,
while the remaining part of the heterostructure remains unchanged.  
This has been explained \cite{TokuraAPL01, NandaPRL07} to be due to the Anderson-Hasegawa-de Gennes double exchange (DEX) interaction \cite{AndersonPR55,ZenerPR51, DeGennes}
 between the interfacial 
Mn magnetic moments and the leaked electrons 
from the metallic CRO side to the CMO side. 
The leaked electrons  occupy the itinerant Mn-$e_g$ states, which then mediate the DEX interaction between the Mn-$t_{2g}$ core moments,  fixed on the lattice sites.
The amount of leaked electrons is sufficiently large to  produce a spin canted state in the interfacial MnO layer,
resulting in a robust net magnetic moment of about $0.85 \mu_B$ per interfacial Mn atom.\cite{TokuraAPL01} 

 \begin{figure}
\includegraphics[scale=0.20]{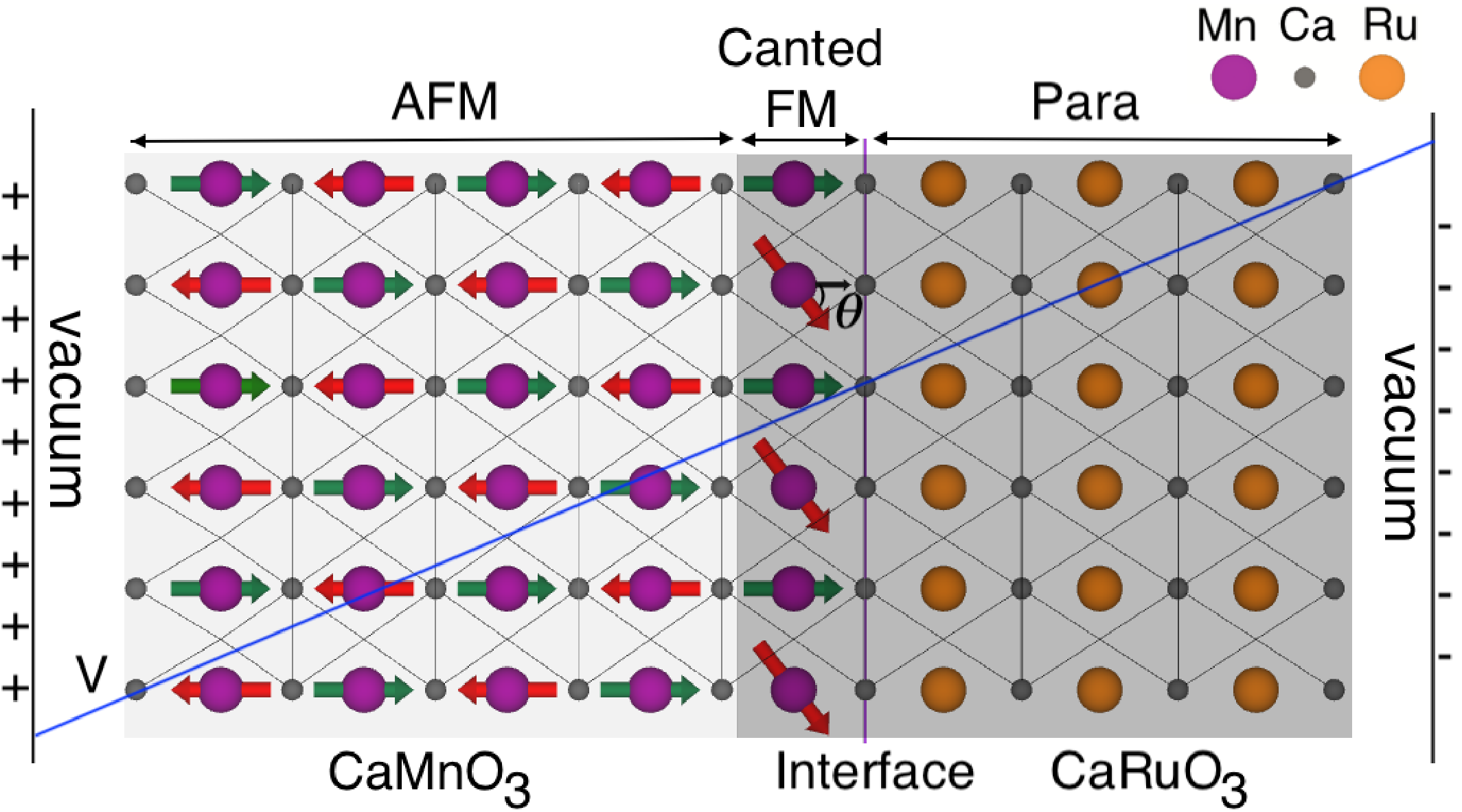}
\caption{Schematic diagram of the CMO/CRO heterostructure and the spin-canted state in the interfacial MnO$_2$ layer. 
Shown is  the supercell
used in the DFT calculations along with the  electric potential seen by the 
electrons (blue line) due to the applied electric field. 
By increasing the charge transfer across the interface, the electric field enhances the 
interfacial ferromagnetism via double exchange by reducing the canting angle $\theta$. 
Apart from the enhanced magnetism of the interfacial MnO$_2$ layer,(001)
we find that
the anti-ferromagnetism of the remaining CMO layers remain more or less unaffected by the electric field.
}
\label{Fig-structure}
\end{figure}

In the DEX mechanism, the spin canting angle is quite sensitive to the itinerant carrier concentration $x$, driving the AFM state into a
spin-canted state at first and eventually into an FM state with increasing $x$.
This is  apparent from the De Gennes expression \cite{DeGennes} for the 
canting angle, to which we return later, viz.,
$\theta_c = 2 \cos^{-1}   (2^{-1}  |t| x / J)$, where $t$ is the  electron hopping integral and $J$ is the AFM Heisenberg exchange.
It is therefore expected that an applied electric field would affect the DEX interaction by modifying the carrier concentration in the
magnetic layers. However, the extent of this effect is unknown since dielectric screening theory indicates merely that the polarization charges would accumulate somewhere in the boundary regions, not necessarily in the magnetic layers. 
 Therefore, this issue needs to be studied in detail. 
Indeed, as our density functional calculations find, much of the surface polarization charges, for example,
 appear in the vacuum region. 
 It is only the carriers that appear in the magnetic layers that matter as far as the DEX mechanism is concerned.

We have chosen the prototypical CRO/CMO system for our work, since there are already several experimental studies on this system reported in the literature.
 In fact, Grutter et al.\cite{Suzuki2015} have recently studied experimentally the electric field dependence of magnetism in this system.
 They find an increase of the ferromagnetic moment with an applied electric field and conclude that it originates from the interface MnO$_2$ layer.


 In this work, we study the 
 effect of an electric field on the electronic structure and magnetism of the CRO/CMO interface in the slab geometry from density-functional calculations.
 We find a text-book like dielectric screening of the applied field,
 which leads to a charge accumulation at the slab surfaces and the interface. 
 However, quite interestingly, not all screening charges occur 
 in the surface or the interface atomic layers.
 For example, the surface polarization charge is found to occur outside the nominal surface, with little or 
 no charge accumulated on the Mn surface layers or the bulk layers. 
As for the interfacial Mn layer, a significant amount of extra charge does accumulate there, which reduces the spin canting angle via double exchange when the electric field is applied, leading to  an increased net ferromagetic moment as a result.

\section {Density-Functional method} 
In our calculations, we considered a slab consisting of five layers of CMO and three layers of CRO,  (CMO)$_5$/(CRO)$_3$,
with each layer consisting of two formula units to  describe the anti-ferromagnetic Mn moments in CMO.
An extra layer of electrically neutral CaO was added as shown in Fig. \ref{Fig-structure},
so that the metal-oxygen octahedra MO$_6$ is complete on both surfaces. 
Test calculations using larger number of layers did not substantially change the results.
We used the same in-plane lattice
constant as the bulk CRO ($a=5.27$ \AA), while the out-of-plane lattice constant was adjusted to conserve the bulk volume of each constituent
material and  a vacuum region of ~$14$~\AA~ was added on each side of the
slab. 
A sawtooth shaped electrostatic potential was added, as indicated by the dashed line in Fig. \ref{Fig-structure},
which was the supercell we used in the DFT calculations. 
Dipole correction was included following the work of Bengtsson. \cite{Dipole}

The atomic positions were relaxed using  the Projector Augmented Wave method (PAW)\cite{PAW, VASP2}
in the generalized gradient approximation (GGA) for the exchange-correlation functional
as implemented in the Vienna Simulation Package (VASP). \cite{VASP1}
The Quantum Espresso  code \cite{espresso1}
was used 
to study the effect of the external electric field on the electronic and magnetic properties,
where 
a norm-conserving ultrasoft pseudo-potential was used together with
the GGA exchange-correlation functional with the Hubbard parameters $U=5 $ eV and $J=0$ 
for the Mn atoms.


\section {Dielectric screening charges: Model and DFT Results} 

\begin{figure}
\includegraphics[scale=0.25]{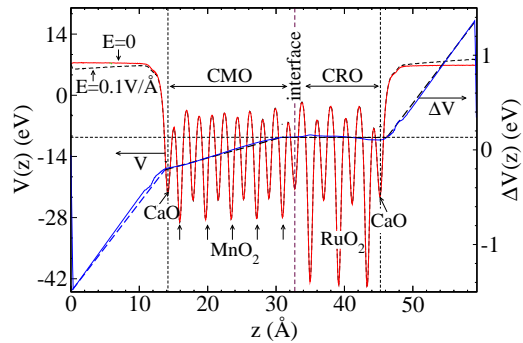}
\caption{Planar averaged  potential $V (z)$ seen by the electron both
with (red line) and without the electric field (black dashed line).
The difference between them $\Delta V$, shown as the blue line,
follows a text-book like  linear behavior in each dielectric region as predicted from the dielectric model. 
The  dashed line next to the blue line is a guide to the eye indicating the piece-wise linear behavior,
the slope of which  
yields the screened electric field.
}
\label{Fig-DFT-Potential}
\end{figure}

\begin{figure}
\includegraphics[scale=0.25] {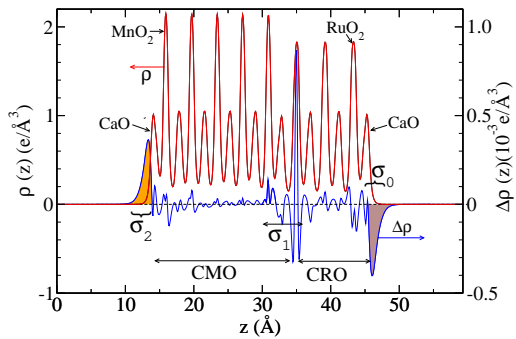}
\caption{Planar averaged electron density  for $E = 0$ (red line) and the extra electrons
accumulated (polarization charge) (blue line)
when the electric field $E = 0.1$ V/ \AA\ is applied. 
The colored areas under the blue line indicate the net accumulation  of charges (positive or negative) 
at the  two surfaces, which are listed in Table \ref{Table1} from direct integration.
 } 
\label{Fig-DFT-Charge}
\end{figure}

The results of our DFT calculations, both with and without an electric field, are shown in Figs. \ref{Fig-DFT-Potential}  and \ref{Fig-DFT-Charge},
where we have shown the planar averaged Kohn-Sham potential $V (z)$ and the charge density $\rho (z)$, respectively.
The planar-averaged quantities are given by the expression $ V (z) = A_{\rm cell}^{-1} \int_{\rm cell} V (\vec r) d^2 r$ and 
similarly  for $\rho (z)$, where the integration is along the plane, normal to the interface, and 
$A_{\rm cell}$ is the surface cell area.
The positions of the individual atomic layers such as MnO$_2$ can be identified in both figures from the $\delta$-function like peaks.
As seen from Fig. \ref{Fig-DFT-Potential}, the
 planar-averaged quantities for with and without the electric field nearly overlap with one another, since the differences are very small.
 The differences,
$\Delta V (z)$ and $\Delta \rho (z)$, induced by the electric field are shown as blue lines in Figs. \ref{Fig-DFT-Potential}  and \ref{Fig-DFT-Charge},
respectively, on an exaggerated scale.

\begin{figure}
\includegraphics[scale=0.25]{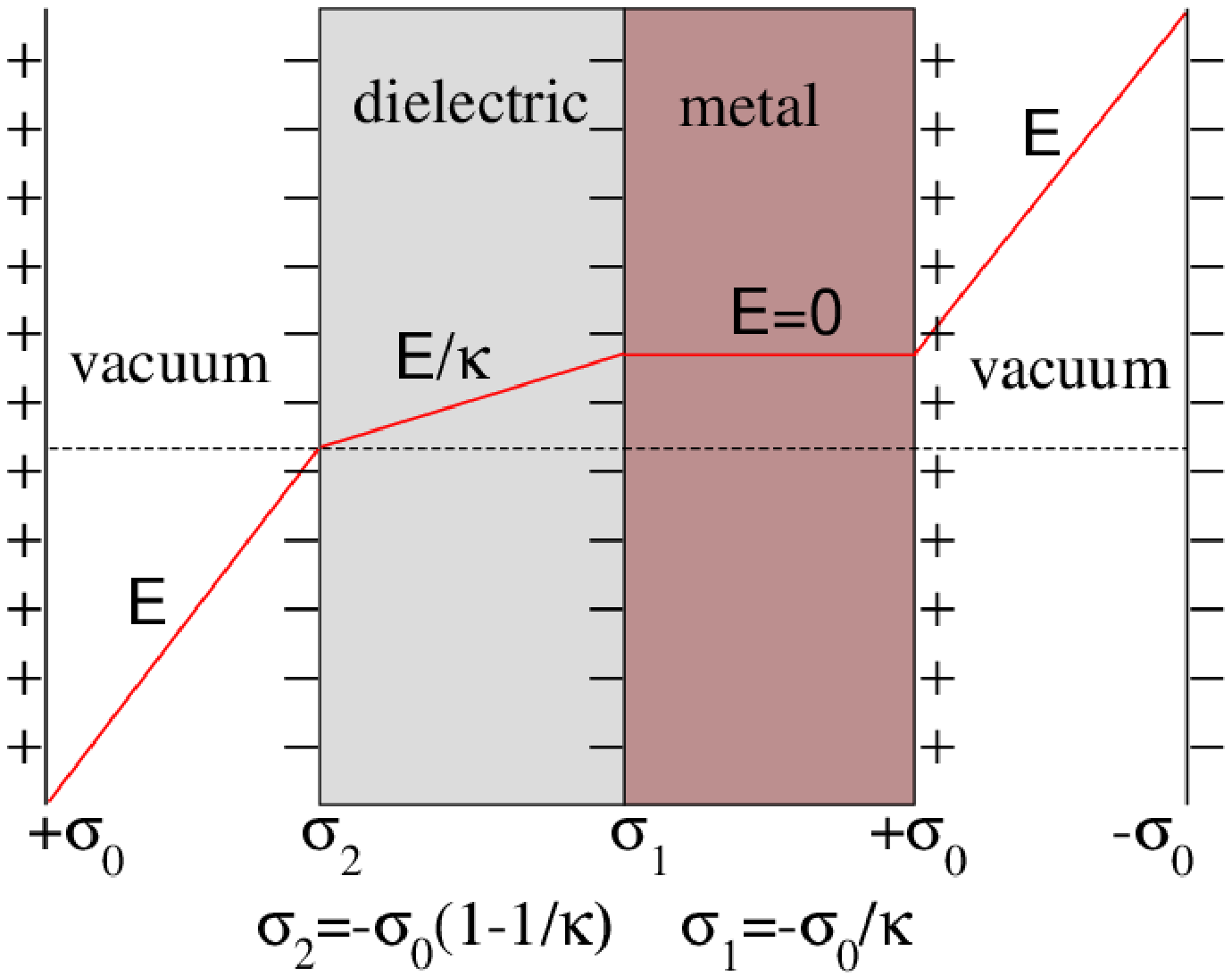}
\caption{Polarization charge accumulated at the boundaries  for the metal/dielectric slab
 from elementary electrostatics, relevant for the CMO/CRO film. 
The slab is placed between two capacitor plates,  $\kappa$ is the 
relative permittivity 
of the insulator, and $\sigma_i$'s indicate the surface charge densities that accumulate at the three
boundaries.
}
\label{Fig-electrostatics}
\end{figure}

The DFT results reveal a remarkable text-book like behavior for the dielectric screening. Points to note are: (a) Piecewise linear potentials in all regions of the slab (Fig. \ref{Fig-DFT-Potential}), corresponding to the screened electric fields predicted by elementary electrostatics theory (Fig. 
\ref{Fig-electrostatics})  and  (b) Accumulation of the screening charges at the two surfaces and the interface layer. 

A somewhat surprising result is that the screening charges at the two surfaces with the vacuum do not occur on the surface atomic layers
as might have been anticipated, 
but they rather occur well inside the vacuum region.
As seen from Fig. \ref{Fig-DFT-Charge}, where the polarization charges at the two surfaces have been indicated by colored areas,
the polarization charges occur outside the surface CaO layers, at a distance of $\sim$1.3 \AA\  away from the atomic planes.

%

%


The screened potential from the DFT calculations
(Fig. \ref{Fig-DFT-Potential})  compares very well with the results of the dielectric screening 
from elementary electrostatics theory, shown in Fig. \ref{Fig-electrostatics}.
The heterostructure is placed between two capacitor plates that produce the electric field $E$.
In the dielectric model, the polarization charges at various boundaries are determined from the Gauss' Law, and these are
indicated in Fig. \ref{Fig-electrostatics}. 
Taking $\kappa$ to be the dielectric constant of the
 insulator (CMO), the surface charge densities are:  $\sigma_0=-\epsilon_0 \text E$
 at the metal surface,
 where $\epsilon_0$ is the vacuum permittivity, 
 $\sigma_1=-\sigma_0/\kappa$ is the charge density at the interface between the 
 metal and the dielectric,
 and $\sigma_2=-\sigma_0(1-1/\kappa)$ is the charge density at the
surface dielectric surface. 
Taking the value $\kappa \approx 5$ to fit with our DFT results for the screening charges
 and the  
vacuum permittivity $\epsilon_0 =  8.85 \times 10^{-12} F/m = 
  5.53\times 10^{-3} |e|/$ (V$\cdot $  \AA), for the case E = 0.1 eV/\AA, we get the numerical values:
 $\sigma_0 = 5.53 $, 
 $\sigma_1= - 1.11$,  and $\sigma_2=-4.42 $,
 in units of $10^{-4}$ $|e|$/\AA$^2$.
 These values together with the corresponding  DFT results have been listed in Table \ref{Table1}.
 
 The DFT values were computed 
 by integrating the planar averaged charge difference $\Delta \rho$ near the CRO and CMO surfaces  indicated by the colored areas  in Fig. \ref{Fig-DFT-Charge}.  
 The computed values are 
 $\sigma_0^{\text DFT}= 5.4 \times 10^{-4}$ $|e|$/\AA$^2$ and $\sigma_2^{\text DFT}=-4.3 \times 10^{-4}$ $|e|$/\AA$^2$.
  Since the interface charge $\sigma_1$ is relatively smaller and charges fluctuate
 quite a bit near the CMO/CRO interface, we were not able to get the value of $\sigma_1$ reliably by direct integration. 
Instead, we obtained $\sigma_1$
from the charge neutrality condition, viz.,
 $ \sum_{i = 1}^3 \sigma_i = 0$, using the integrated values for $\sigma_0$ and $\sigma_2$,
 with the result $\sigma_1^{\text DFT}=-1.1 \times 10^{-4}$ $|e|$/\AA$^2$. 
 All these values agreed quite well with the polarization charges obtained from the dielectric model (Table \ref{Table1}), 
 assuming the dielectric constant to be $\kappa \approx 5$. 
In comparison to this, the corresponding experimental value $\kappa  \approx 7$,  inferred from the optical conductivity data\cite{LoshkarevaPRB04},
 is somewhat larger.
The reason for this difference could be due to the approximate nature of the functionals used in the DFT calculations 
or due to the small number of layers in the
supercell used, so that the bulk dielectric screening limit has not been reached.

\begin{table} [bt]
\caption{Surface polarization charge densities induced by  the applied electric field at the interface ($\sigma_1$) and the two surfaces ($\sigma_0$ and $\sigma_2$), computed from the DFT as well as from the electrostatics theory. The applied electric field  is $E = 0.1 $  V/\AA, the relative permittivity
 $\kappa = 5$ is used in the dielectric model, and the surface charge densities are expressed in units of $10^{-4}  |e|$/\AA$^2$.
}
\begin{ruledtabular}
\begin{tabular}{ c|c|c|c}
  &$\sigma_0$ & $\sigma_1$ & $\sigma_2$ \\  
\hline 
DFT&5.4  & -1.1 & -4.3\\ 
Dielectric model& 5.53 & -1.11 & -4.42 \\
\end{tabular}
\label{Table1}
\end{ruledtabular}
\end{table}
 
 As seen from  Figs. \ref{Fig-DFT-Potential} and \ref{Fig-electrostatics} and Table \ref{Table1}, the DFT results agree quite well with the text-book like screening profile
 including the screened electric fields and the polarization charges at the boundaries.
 Fig. \ref{Fig-DFT-Potential} shows  that the final screened electric fields in various regions are uniform (linear $\Delta$V)
 as expected from the electrostatics model.
 While in the vacuum region, the applied electric field is unchanged, it is completely screened in the metallic region (CRO) 
 as expected ($\kappa = \infty$)
 and is reduced by the dielectric constant $\kappa$ in the insulating region (CMO).
 Taking the ratio of the  screened electric field in the CMO region  to the applied electric field (Fig. \ref{Fig-DFT-Potential}),
we get a second estimate  $\kappa \approx 4.4$, which is similar to the value $\kappa \approx 5$ obtained from the surface polarization charges
discussed above.

\begin{figure} [bt]
\includegraphics[scale=0.425]{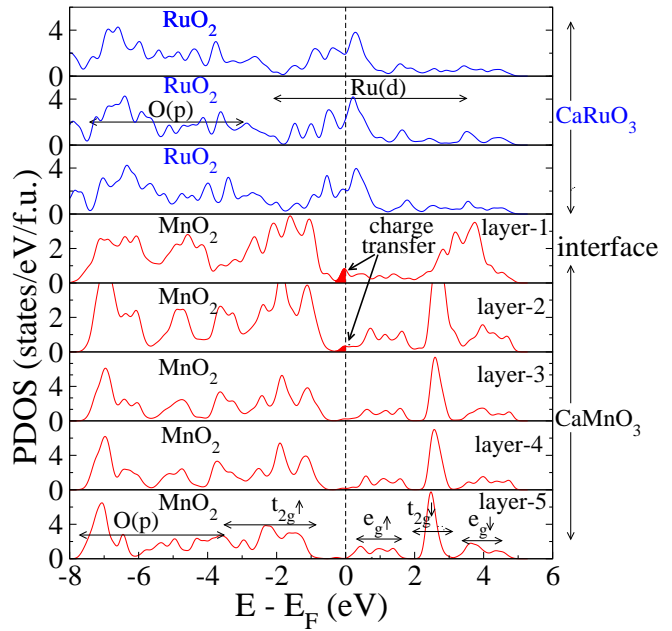}
\caption{Partial densities of states (PDOS) for Mn and Ru layers in units of states/eV/ formula unit (MnO$_2$ or RuO$_2$)
including both spins. 
Charge transfer across the interface to the CMO side (primarily Mn-e$_g$ states) are indicated by two arrows
in the middle two panels.
Here, the electric field is  $E = 0$. With the applied electric field, the figure remains more or less the same except that the charge transfer
 to the CMO side is a bit larger as listed in Table \ref{Table2}.
}
\label{Fig-DFT-PDOS}
\end{figure}

As already mentioned, we find that the polarization charges do not necessarily reside on the atomic layers.
For our purpose, it is important to study the electronic charges on the individual atomic layers, especially the Mn layers,
as the itinerant Mn-e$_g$ electrons mediate the DEX between the core t$_{2g}$ spins leading to spin canting.
For this purpose, we have computed the layer-resolved partial density of states (PDOS) on the individual MnO$_2$ and RuO$_2$ layers, 
which are shown in Fig. \ref{Fig-DFT-PDOS}.
In the CMO bulk, the material is an insulator with filled majority-spin t$_{2g}$ bands and empty e$_g$ bands,
as indicated in the bottom panel of Fig. \ref{Fig-DFT-PDOS}. 
There is some charge transfer across the interface from the RuO side to the two neighboring MnO$_2$ layers
as indicated in the figure. 
By directly integrating the area of the occupied Mn-e$_g$ states (marked in red in Fig. \ref{Fig-DFT-PDOS}), we can compute the   
charge transfer into various MnO$_2$ layers in the structure. There is significant charge transfer only to the first two MnO$_2$ layers at the interface as indicated in 
Fig. \ref{Fig-DFT-PDOS}.

The charge transfer to the  various MnO$_2$ layers from the RuO side are also listed in  Table \ref{Table2}. 
Without  the electric field, there is already a charge transfer from the CRO side to the CMO side\cite{NandaPRL07}. 
This leads to a net dipole moment with a positive charge on the CRO side and a negative charge on the CMO side, but there is no net monopole charge.
As seen from Table \ref{Table2}, for $E = 0$, the charge accumulated on the first MnO$_2$ layer is 0.117 $e^-$/Mn atom $\approx$ 8.4 $\times 10^{-3}$  $e^-$/\AA$^2$. 
The accumulated electrons occupy the Mn $e_g$ states, serving as the itinerant electrons that mediate the double exchange between the Mn $t_{2g}$ core spins, which we discuss in more detail in Section \ref{Spin}.

When the electric field is applied, there are monopole charges $\sigma_i$ that accumulate at various boundaries in
order to screen out the applied  field. These add to the layer charges already 
existing for $E = 0$.
Table \ref{Table2} shows that with $E = 0.1 $  V/\AA, the first MnO$_2$ layer gains a
small  additional charge making the total in that layer to be 0.121 $e^-$/Mn atom, 
which translates into an additional charge of 0.004 $e^-$/Mn atom ( -2.9 $\times 10^{-4}$ $|e|$/\AA$^2$). 
Note that although it is of the same order of magnitude as the  interface polarization charge $\sigma_1$ seen from Table \ref{Table2},
they are not necessarily the same, as  $\sigma_1$ is not necessarily located entirely on the interfacial MnO$_2$ layer. 
For the DEX interaction on the interfacial MnO layer, it is only the net charge (itinerant $e_g$
electrons) on that layer that matters, not the total polarization charge that accumulates in the interface region due to the dielectric screening.
As seen from Fig. \ref{Fig-DFT-Charge}, the polarization charge $\sigma_1$ is spread over several monolayers
at the interface region, both on the CMO and the CRO sides.

As we move away from the interface, the accumulated charge in the MnO$_2$ layers quickly  reverts to the bulk value
as seen from Table \ref{Table2}. 
The bulk limit is already reached as quickly as the third layer and beyond.  
An interesting point to note regarding the surface charges at the vacuum interface is that even though there is a considerable polarization charge
($\sigma_2 = - 4.3 \times 10^{-4} |e|/$\AA$^2$ at the CMO/vacuum interface
from Table \ref{Table1}), only a small fraction of it appears on the surface MnO$_2$ layer
(layer-5 in Table \ref{Table2}). Indeed as observed already, much of the charge of both $\sigma_2$ and $\sigma_0$ at the two surfaces appear well inside the vacuum region,
with the peaks appearing about $1.3$ \AA\ outside of the terminal CaO surface.
Thus, the surface MnO$_2$ layer being more or less similar to the bulk, with very little additional charge transfer due to the electric field, the magnetism continues to remain anti-ferromagnetic,
i. e., the same as in the bulk. 
This is also confirmed from the total energy calculations within the DFT,
which shows the MnO$_2$ surface layer to remain anti-ferromagnetic.


\begin{table}
\caption{Extra  electrons per Mn atom,
as compared to the bulk, accumulated at various MnO$_2$ layers near the interface.
The electrons occupy the Mn $e_g$ states as indicated from Fig. \ref{Fig-DFT-PDOS}.
Electric field $E$ is in units of  (V/\AA). These numbers are to be multiplied with the factor $7.2 \times 10^{-2}$ to get the electron numbers in units of $e^-/ $\AA $^2$ for the corresponding MnO layer for comparison with the polarization charges shown in Table \ref{Table1}.
}
\begin{ruledtabular}
\begin{tabular}{ c|c|c|c|c|c}
E&layer-1 & layer-2 & layer-3 & layer-4 & layer-5\\  
\hline
0&0.117 & 0.044 &0.002 &0.000 &0.000\\ 
0.1&0.121 &0.045 & 0.002& 0.001&0.002\\ 
\end{tabular}
\label{Table2}
\end{ruledtabular}
\end{table}



\section {Spin canting and interfacial ferromagnetism }
\label{Spin}

\subsection { Ferromagnetism at the interface} 
To study the stability of the interfacial magnetism, we computed the total energy of the two magnetic structures,
one where all Mn atoms are anti-ferromagnetic (AFM) as in the bulk and a second structure, where only the
interfacial MnO$_2$ layer is ferromagnetic (FM), while the remaining layers retain the AFM structure of the bulk.
We find that even though 
 the energy difference between the FM and AFM configurations for the Mn-Mn bond at the interface is 
somewhat sensitive to the magnitude of the Coulomb repulsion parameter $U$
used in the DFT calculations, the FM state is always more stable. 
 In  Table  \ref{Table3}, we have listed the results for $U = 2$ eV, 
for which the computed  
$\Delta E \equiv E_{\uparrow \downarrow} - E_{\uparrow \uparrow}  = 16.1 $ meV value
for $ E =0$ is comparable to the experimental value of 13.1 meV for the bulk CMO.\cite{Wollan, Konig} 
The other quantities such as the charge transfer and the density-of-states are not sensitive to the value 
of $U$, and were calculated with $U = 5$ eV.

Table  \ref{Table3}  shows that
the FM state of the interfacial MnO$_2$ layer is more stable both with and without  the external electric field.
 As discussed later, the spin canted state, which has a reduced net FM moment,
  has actually even lower energy than the FM state,
 which has been confirmed earlier for the intrinsic sample ($E = 0$) both from experiment and theory.\cite{TokuraAPL01, NandaPRL07}
With the application of the electric field,  
the total energy of the FM state is further reduced by about 3 meV,
making the FM state even more stable in the presence of an electric field. 
As already mentioned, we have also computed the same energy difference for the surface MnO$_2$ layer 
and find that, in contrast to the interfacial MnO$_2$ layer,
the AFM state at the surface continues to remain energetically favorable, both with and without the electric field.

These results are consistent with the Anderson-Hasegawa DEX result, 
that the FM state becomes progressively more energetically favored over the AFM state as the itinerant carrier concentration
is increased, in our case by the application of the electric field.
However, the lowest energy state is neither FM nor AFM, but a spin canted state, 
and we discuss this by considering a simple DEX model on a square lattice, that describes the 
magnetism of the interfacial MnO$_2$ layer.


\begin{table} 
\caption{Calculated total energy, where the CMO layer at the interface is either 
FM or AFM, with the remaining layers being AFM, i.e., the same as in the bulk.  
Energies are per interfacial bond and in units of meV for the electric fields $E = 0$ and $E =  0.1 $ V/\AA. 
} 
\begin{ruledtabular}
\begin{tabular}{ c|c|c}
E&FM & AFM\\
\hline
0&-16.1 & 0\\
0.1&-19.2 &0\\
\end{tabular}
\end{ruledtabular}
\label{Table3}
\end{table}

\subsection { Double-exchange model and spin canting}

We consider the well known Anderson-Hasegawa double exchange model 
\cite{AndersonPR55,ZenerPR51,DeGennes,MishraPRB97}
and apply it to a square lattice appropriate for the MnO$_2$ layer.
The Hamiltonian is
\begin{equation}
{\cal H}  =t \sum_{\langle   ij   \rangle \sigma} c_{i\sigma}^{\dagger}c_{j\sigma}+ h. c. 
+\sum_{\langle    ij  \rangle}J {\hat  S}_i. {\hat  S}_j    -     2J_H\sum_{i}{\vec S}_i.{\vec s}_i,
\label{Eq-AH}
\end{equation}
which describes the motion of  the itinerant Mn $(e_g$) electrons (the  corresponding field operators
are $c_{i\sigma}^{\dagger},c_{j\sigma}$ with  $i$ and $\sigma$ being the site and the spin indices)
moving in a lattice of Mn $t_{2g}$ core spins ($S = 3/2$).
Here  $t$ is the tight binding nearest neighbor hopping, 
${\bf s}_i=1/2\sum_{\mu \nu} c_{j\mu}^{\dagger}{\bf \tau}_{\mu\nu}c_{j\nu}$ 
is the spin of the 
itinerant electron, with
 the Pauli matrices $ 
{\bf \tau}$,  $J$ is the superexchange, $J_H$ is the Hund's coupling,
and the angular brackets indicate sum over distinct pairs of bonds in the lattice. 
Typical parameters for CMO are\cite{NandaPRL07, Satpathy}: $t =- 0.15$ eV, $J = 7 $ meV, and $J_H = 0.85 $ eV.

It is instructive to consider the de Gennes result\cite{DeGennes}  for the limiting case $J_H = \infty$,
which suggests a spin-canted state in the presence of the itinerant carriers.
In this limit, since only one spin channel parallel to the core spins
is available for the itinerant electrons, Eq. (\ref{Eq-AH})  is equivalent to
the spinless Hamiltonian 
${\cal H}  = \sum_{\langle   ij   \rangle }   t \cos (\theta_{ij} / 2) \ c_i^{\dagger}c_j+ h. c. 
+\sum_{\langle    ij  \rangle}J {\hat  S}_i. {\hat  S}_j  $, 
where the  hopping has been modified by the well known Anderson cosine factor\cite{AndersonPR55}, 
with $\theta_{ij}$ being the polar angle
difference between the neighboring core spins.
Taking a bipartite square lattice, with the spins in the two sublattices (A and B) canted by the angle $\theta$ with respect to one another,
and considering a small concentration $x$ of the itinerant carriers, 
 the electrons occupy the band bottom $E_b = -z |t| \cos (\theta/2)$, where $z = 4$ is the number of nearest neighbors. 
 The canting angle $\theta_c$ is obtained by minimizing the total energy
\begin{equation}   \label{Energy}
E = E_b x + (z/2) J \cos \theta,
\end{equation}
which  yields the  result
\begin{equation}  \label{angle}
\theta_c = 2 \cos^{-1}   \big( \frac{ |t| x} {2 J}  \big) . 
\end{equation}

 For the $J_H = $ finite case, no such analytical result is possible, and we must solve for the band structure energies
 $\varepsilon_{nk}$ by keeping both spin channels in the Hamiltonian (\ref{Eq-AH}) and sum over the occupied states.
 The canting angle $\theta_c$ is obtained by numerical minimization of the total energy
 \begin{equation} \label{band-energy}
E=(z/2) J \cos \theta +        \sum_{n{ k}}^{\rm  occ}   \varepsilon_{n{ k}}.
\end{equation}

\begin{figure} 
\includegraphics[scale=0.35]{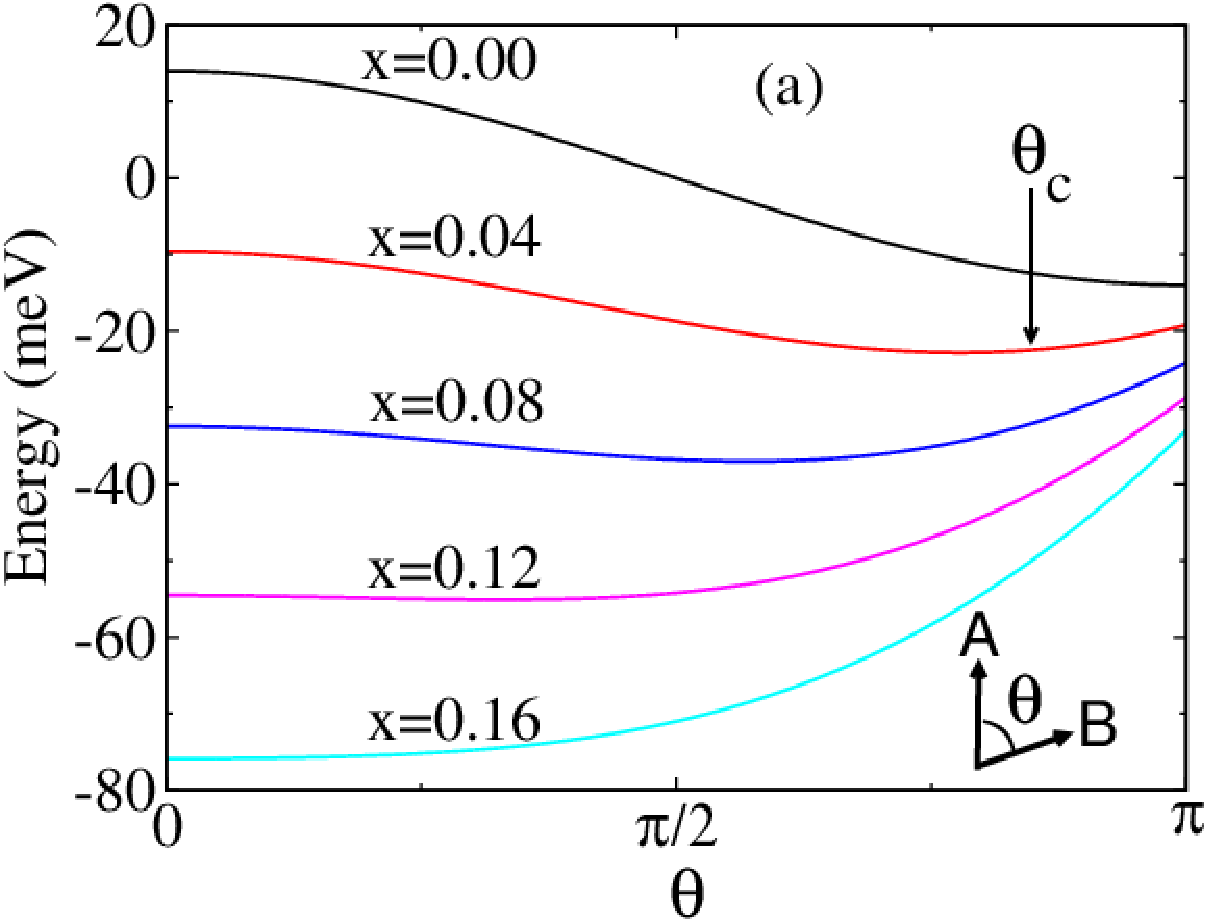}\\
\vspace{8mm}
\hspace*{-0.25cm}\includegraphics[scale=0.29]{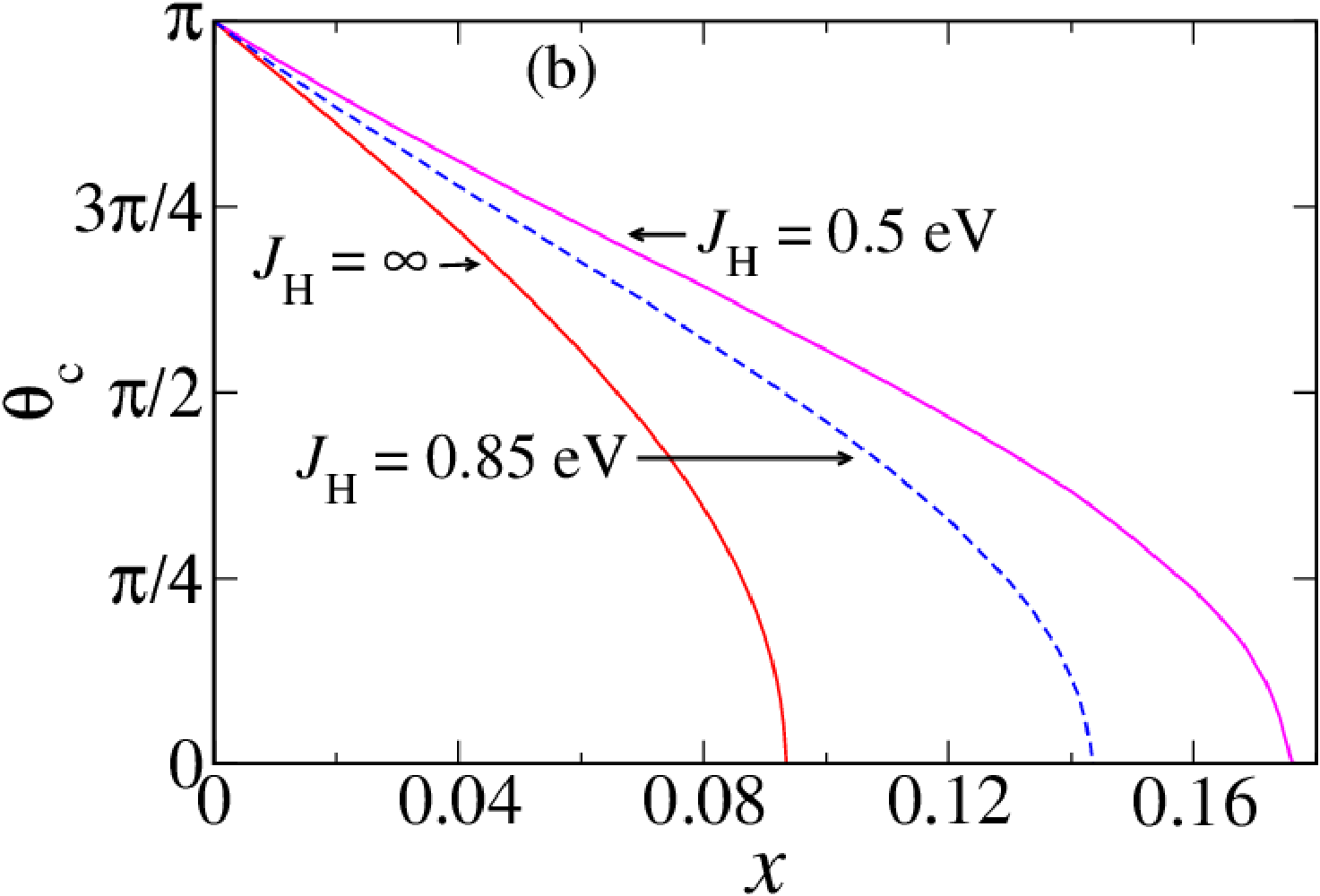}
\caption{Energetics of the spin canted state. (a) Energy from Eq. (\ref{band-energy})  
as a function of the  angle  
$\theta$ between spins in the two sublattices, A and B, for   several values of  the electron concentration $x$.
The minimum yields the canting angle $\theta_c$ (indicated by an arrow for the $x = 0.04$ case).
Starting from  the AFM state ($\theta_c = \pi$) for $x = 0$, the system turns into an FM state ($\theta_c = 0$) beyond the 
critical concentration $x_c \approx 0.135$, so that one obtains an FM state for the case $ x = 0.16$.
The parameters are: $J = 7$ meV, $t = - 0.15$ eV, and $J_H = 0.85$ eV.
(b) The spin canting angle $\theta_c$ as a function $x$ for three cases: $J_H = 0.5$ eV, $ 0.85$ eV, and $ \infty$,
with the parameters $J$ and $t$ being the same as in Fig. (a).
The critical concentration $x_c$, beyond which an FM state is obtained ($\theta_c = 0$),  
is where the curves meet the $x$ axis. 
With increasing $x$, $\theta_c$ decreases, leading to an enhancement of the net ferromagnetic moment.
}
\label{Fig-Canting}
\end{figure}

The computed total energy is shown in Fig. \ref{Fig-Canting} (a) as a function of the canting angle $\theta$ for various electron  concentration $x$.
As seen from the figure, when the electron concentration  $x = 0$, the minimum energy 
occurs at the canting angle $\theta_c = \pi $ resulting in an AFM state, obviously due to the super exchange interaction $J$, 
which is the only interaction without any itinerant carriers.
With increasing $x$, the strength of the DEX interaction slowly increases, producing a spin canted state, and eventually, beyond a critical value  $ x > x_c$,  the DEX dominates resulting in an FM state ($\theta_c = 0$).

 The critical concentration in the $J_H = \infty$ limit is given by Eq. (\ref{angle}) 
 and has the value 
 $x_c  = 2 J / |t| \approx 0.09\  |e|$/ interfacial Mn atom, which is also seen from Fig. \ref{Fig-Canting} (b),
 where we have presented the concentration dependence of the canting angle for several values of $J_H$. 
 It is clear that for  $J_H = 0$, the itinerant and the core spins are not coupled and therefore the system remains AFM for all $x$,
due to the super exchange interaction of the core spins, up to the full occupation of the bands. 
As $J_H$ is increased from zero, the critical concentration $x_c$ monotonically decreases,
 eventually approaching the 
de Gennes result $x_c  = 2 J / |t|$ for $J_H = \infty$.
The critical values of $x_c$ shown in Fig. \ref{Fig-Canting} (b)  for the three values of $J_H$ are consistent with this expectation.

We now discuss the effect of the electric field on the charge transfer across the interface into the MnO$_2$ layer, which in turn 
affects the spin canting and therefore the net ferromagnetism.
As seen from Table \ref{Table2}, there is already a significant charge leakage to the interfacial MnO$_2$ layer
even for $E = 0$, which leads to a canted AFM state.
The magnitude of the canting angle $\theta_c$ can be estimated from 
Fig. \ref{Fig-Canting} (b).
With the applied electric field, the charge transfer increases  due to the build up of the dielectric screening charges.
 As a result, the canting angle decreases, thereby leading to the enhancement of the net FM moment.
 The net FM moment per Mn atom in the MnO$_2$ layer is given by the expression $ m = m_s (1 + \cos \theta_c)/2$,
 where $m_s \approx 3 \ \mu_B$ is the Mn core spin moment.
If we take $J_H \approx 0.85 $ eV for CMO\cite{Satpathy}, 
the predicted increase obtained from Fig. \ref{Fig-Canting} (b) is from  $ m = 2.1\  \mu_B$
($E = 0$) to 2.3 $\mu_B$
($E = 0.1$ V/\AA), corresponding to the change in the electron concentration of $x = 0.117$ to 0.12 electron/ Mn atom,
as seen from Table \ref{Table2}.  

Indeed, such an enhancement of the net FM moment has been observed in the neutron
reflectivity experiments\cite{Suzuki2015}. 
However, the experiment shows a much larger increase
in the FM moment, viz.,  from 1 $\mu_B$ to 2.5 - 3.0 $\mu_B$, corresponding to the transition from a canted AFM state to a fully FM state of the Mn$^{+4}$ ion at the interface.
However, notice from Fig. \ref{Fig-Canting} (b) that the canting angle is quite sensitive to the
itinerant carrier concentration $x$ in the interfacial MnO$_2$ layer, and a critical
value of $x_c \approx 0.14$ (for $J_H = 0.85$ eV) would turn the system
completely ferromagnetic.
This reflects an increase of the itinerant carriers by just 0.02 $|e|/$ Mn atom by the
electric field on top of the $ \sim 0.12 |e|/$ Mn atom that already exists in the intrinsic 
interface for $ E =0$.

Even though the theory and experiments agree qualitatively on the 
increase of the FM moment with the electric field, a quantitative comparison is difficult owing to several factors. 
One, it is difficult to experimentally determine
the exact magnitude of the electric field that is applied to the CRO/CMO heterostructure, since the structure is capped 
by several other layers of materials in the actual sample.\cite{Suzuki}
Second, transition-metal oxide samples are notorious for the oxygen stoichiometry issues and 
it is quite conceivable that the applied electric field leads to a migration of the oxygen atoms to the interface, leading to
an extra mechanism of charge accumulation at the interface. 
Since the double exchange mechanism becomes stronger with an increase of the carrier concentration $x$,
this would increase the tendency towards ferro-magnetism,
and as already pointed out just an extra 0.02 $|e|/$ Mn atom at the interface
is needed to drive the system completely ferro-magnetic.
Finally, there may be substrate-induced strain in the interface, which 
was not studied in the experiment, nor was it considered in our theory.
It would be desirable to study these effects further.

\section{Summary}
In summary, we studied the effect of an external electric field on the CRO/CMO (001) interface using density functional methods
in order to understand the field tuning of the magnetism at the interface.
This system was chosen due to the existing experiments, but the conclusions should be valid for a variety of interfaces.

We found several interesting results.  
(1) The polarization charges induced  at the interface and the surfaces with the vacuum to screen the applied electric field
followed a text-book like profile. 
(2) Interestingly, the surface polarization charges occurred well inside the vacuum (at a  distance of about 1.3 \AA\ from the 
surface atomic planes). Similarly, the interface polarization charge is spread over several atomic planes in the interface region, 
which means that not necessarily all of it participate in the interface phenomena such as the double exchange in our case. 
(3) The surface MnO$_2$ layer is predicted to remain AFM as in the bulk, so that the enhancement 
in the ferromagnetism seen in the  experiments is unlikely to come from the surface, as has been suggested in the experiments.\cite{Suzuki2015} 
(4) Our theoretical work  supports the experimental observation that the interfacial magnetism is enhanced by the applied field and 
identifies the
extra charge accumulation at the interface MnO$_2$ layer and the double exchange mechanism to be responsible
for the enhancement. However, the effect is much stronger experimentally than the theory predicts. The difficulty of a quantitative comparison with the experiment
is due to several factors, viz., 
(i) The possibility of electric-field driven oxygen migration to the interface, 
(ii) Unknown magnitude of the electric field at the interface due to the presence of the substrate and the cap layers in the experiments,
and (iii) Possible strain in the structure due to the substrate.
Nevertheless, both theory and experiment indicate a strong electric field tuning of the interfacial magnetism, 
with potential for application in magnetoelectric devices.


{\it Acknowledgment-- } We thank Professor Yuri Suzuki for stimulating this work and for her insightful discussions. 
We acknowledge financial support from the US Department of Energy, Office of Basic Energy Sciences, Division of Materials Sciences and Engineering Grant No. DEFG02-00ER45818.
Computational resources
were provided by the National Energy Research Scientific
Computing Center, a user facility also supported by the US
Department of Energy.

\bibliography{library}

\end{document}